# Assessing Educational Research – An Information Service for Monitoring a Heterogeneous Research Field

**Karima Haddou ou Moussa**
Department of Knowledge Technologies for the Social Sciences, GESIS Leibniz Institute for the Social Sciences, Unter Sachsenhausen 6-8, 50667 Cologne, Germany. Email: karima.haddououmoussa@gesis.org[1]

**Ute Sondergeld**
Information Center for Education, German Institute for International Educational Research (DIPF), Schloßstraße 29, 60486 Frankfurt am Main, Germany. Email: Sondergeld@dipf.de

**Philipp Mayr**
Department of Knowledge Technologies for the Social Sciences, GESIS Leibniz Institute for the Social Sciences, Unter Sachsenhausen 6-8, 50667 Cologne, Germany. Email: Philipp.Mayr@gesis.org

**Peter Mutschke**
Department of Knowledge Technologies for the Social Sciences, GESIS Leibniz Institute for the Social Sciences, Unter Sachsenhausen 6-8, 50667 Cologne, Germany. Email: Peter.Mutschke@gesis.org

**Marc Rittberger**
Information Center for Education, German Institute for International Educational Research (DIPF), Schloßstraße 29, 60486 Frankfurt am Main, Germany. Email: rittberger@dipf.de

---

[1] Corresponding author

**Abstract**
**The paper presents a web prototype that visualises different characteristics of research projects in the heterogeneous domain of educational research. The concept of the application derives from the project "Monitoring Educational Research" (MoBi) that aims at identifying and implementing indicators that adequately describe structural properties and dynamics of the research field. The prototype enables users to visualise data regarding different indicators, e.g. "research activity", "funding", "qualification project", "disciplinary area". Since the application is based on Semantic MediaWiki technology it furthermore provides an easily accessible opportunity to collaboratively work on a database of research projects. Users can jointly and in a semantically controlled way enter metadata on research projects which are the basis for the computation and visualisation of indicators.**

## Background

In recent years, educational research has been focused by social and political discourse. On the one hand, this awareness derives from educational policy objectives jointly agreed by member states of the European Union, e.g. Bologna Process, Lisbon Strategy (Lifelong Learning, harmonisation of Higher Education Area, exchange in Vocational Education and Training). On the other hand, large-scale international student assessments have contributed to the development, e.g. TIMSS (Trends in International Mathematics and Science Study) or PISA (Programme for International Student Assessment). Germany has regularly participated in such studies since 1995, and findings from the international comparisons revealed deficiencies of education systems. A need to strengthen research on education was consequently identified to gain evidence for the improvement of education. As a result, educational research in Germany has recently received more attention than it had "for the past 35 years", (Tillmann, 2006).

Educational research is characterised by its multidisciplinary nature. Besides traditional core disciplines of educational science, psychology and social sciences, the field encompasses subject didactics as well as many other disciplines concerned with investigating education systems. Heterogeneity of the field has evolved in consequence of the diversity of different disciplines and science theoretical principles, methods, structures and types of communication. The entire scope of humanities, social sciences and even natural sciences approaches is involved.

Against this background, the project "Monitoring Educational Research" (Monitoring Bildungsforschung (MoBi))[2] targets the analysis of research projects and publications in educational research since the mid-1990s with the aim to develop indicators that highlight structures, developments and types of communication in educational research. The project focuses on assessment of research projects stored in the SOFISwiki[i] database edited by GESIS, where research projects from different social sciences disciplines are systematically recorded, thus, providing a good means of exploring the broad field of educational research. Project outcomes were applied to conceptualise a web-based prototype that provides a visualization of indicators and allows users to run a visually enhanced monitoring of properties and dynamics of a field under study. The paper presents a description of some indicators used for the analysis of the research field (chapter 2). Moreover, technologies and methods applied to the implementation of the developed monitoring prototype are described (chapter 3).

## The role of indicators in scientific research

In science, assessment of developmental processes is generally based on indicators which present reality in terms of numerical relations (Hornbostel, 1999). Indicators can achieve different levels of complexity ranging from simple figures to relative numbers and complex indices (Meyer, 2004). Input figures such as material equipment or human resources are correlated with measurable outcomes, e.g. prizes, publications, doctoral degrees, stipends, informing on activity, structure and quality of a field of research (Hornbostel, 1999). Several factors bear an impact on the validity of indicators: type and scope of available data, research approaches and characteristics of the matter under investigation. Depending on the approach taken, the number of funded projects might serve as an indicator for research achievements, such as success in competitively acquiring funding. From another perspective, external funding can be interpreted as simple input of financial resources. Moreover, assessment of an external funding indicator needs to consider in how far the acquisition of external funding is common to a research discipline: Block, Hornbostel and Neidhardt (1992) have demonstrated that external funding is far more wide-spread in natural sciences than in social sciences, hence, external funding has a different meaning in the disciplines, which should be reflected in a comparison of research domains. The relevance of indicators is furthermore affected by characteristics within the disciplines. Hornbostel (2001) characterises educational science as a discipline that is comprised of humanities, social-scientific and empirical traditions and a part specialised in delivering practical services. In each of these parts within the discipline, a particular indicator plays a different role and it bears a different meaning.

---

2 The official project title is "Entwicklung und Veränderungsdynamik eines heterogenen sozialwissenschaftlichen Feldes am Beispiel der Bildungsforschung". It was funded by the Leibniz Association, subject to the SAW procedure (SAW-2011-DIPF-3), from May 2011 to July 2014. The following institutions have collaborated in the project: GESIS – Leibniz Institute for Social Sciences; Leibniz Centre for Psychological Information and Documentation (ZPID); Institute for Research Information and Quality Assurance (iFQ); German Institute for International Educational Research (DIPF).
http://www.dipf.de/de/forschung/projekte/monitoring-bildungsforschung-mobi

To analyse the research projects we selected such indicators that cover the structure as well as the content of a research project. Existing data did not allow for construction of complex indicators. Against this background, we perceive indicators as metadata that according to their respective character describe different features of a field of research. "Research activity" (Forschungsaktivität) models the development of a field of research as a basic indicator. Taking into account that since the 1990s research funding is predominantly governed by external sources (Schubert & Schmoch, 2010) and, thus, the acquisition of research funding is increasingly gaining importance, the indicator for "research funding" (Förderart) reflects the development in educational research. Development regarding obtainment of degrees, subsumed in the indicator "qualification" (Qualifizierungsarbeiten) demonstrates the state of training for academic research which is highly relevant for the continuity of a discipline and plays a pivotal role in strategies for strengthening educational research (Hauss et al., 2012). The indicator "disciplinary area" (Disziplinbereich) models the subject discipline a project is assigned to, it serves to ascertain what disciplines are active in educational research and reflects the diversity of access to the field. Beyond these indicators, for which an implementation in the web prototype is exemplified below, we examined other indicators such as cooperation, research methods and objectives, biographical aspects and target groups.

## Monitoring Prototype

The aim of the monitoring prototype is to visually present indicators of the development of educational research and to offer a tool that informs about the changes of a research field over time.

*Database and technical background*

Technologically, the monitoring prototype is based on the online platform SOFISwiki. This community platform enables storage and search on social sciences research projects from different fields such as education sciences, psychology, political sciences, from German-speaking countries in Europe (Germany, Austria, Switzerland: D-A-CH). So far, SOFISwiki contains 53,702 project records. The monitoring prototype uses a subset of 9,122 records out of SOFISwiki which contains only completed educational research projects dating from 1995 to 2009.

SOFISwiki is based on the Semantic MediaWiki (SMW)[ii] technology. SMW is a version of MediaWiki, extended by semantic technologies of the platform used by many Wiki applications such as Wikipedia. The purpose of this extension is to enable quick semantic search and discovery of data in a Wiki system (Krötzsch et al., 2007). Therefore, not only pure text pages are stored in a Wiki page, but pages enriched with additional information. These so-called attributes describe the relationship between Wiki pages. Hyperlinks are used to create direct connections between these pages. The page relation is realised either by typed references and/or by values of the attributes. Page names in a MediaWiki system consist of a namespace and a selected name. Namespaces are structuring concepts that are used to group pages. MediaWiki has, for instance, the following namespaces: category, attribute and template. Category[iii] allows the classification of pages. A page can be assigned to one or more categories. The assignment of a page to a category is effected by the following syntax: *[[Category: Category name]]*. The Wikitext *[[Category: MoBi]]* indicates for example that MoBi (Monitoring Educational Research) belongs to the Namespace "Category" and, hence, that "MoBi" is the name of that category. All MoBi-Projects are assigned to the category "MoBi and Projects". They will thus appear at the end of the page of each project as follows: *Categories: MoBi | Projects*. Attributes[iv] are treated as categories for values in Wiki pages, by which semantic data are grasped. The users are allowed to create attributes themselves following this simple scheme: *[[attribute name: attribute value]]*. This Wiki syntax assigns the given attribute "attribute name" the value "attribute value" and displays this value in the respective location on the page: e.g. [[year::1997]]. Using these attributes, a lot of information about the single pages can be explicitly displayed on semantic Wiki pages. They can be used for various kinds of data such as numbers, dates or geographical coordinates where each attribute is assigned a data type; otherwise annotations in unfitting types will simply be ignored. For attribute values, many different data types exist, e.g. String, Page, Number. The property "persons" of a research project is for example an attribute of the type String. Each Wiki page has a list of various attributes and their values, which is referred to as metadata schema. Two types of representation exist for this schema, i.e. user and developer view. Figures 1 and 2 represent screenshots of the attribute list of a Social Science project. The user view (Figure 1) only displays the most important information that is used to describe a project and that is relevant for the user. These include the metadata of the project (title, author, year, etc.), the abstract, and the institutions and research institutions involved. If the project is funded, the sponsor is displayed, too. In addition, the methods used in the project (empirically, empirically-quality, etc.) and tags are shown.

Figure 1: User view in SOFISwiki, displaying metadata on research projects

The developer view (Figure 2) instead contains a lot of information presented as clickable search icons that enable quick discovery of pages with identical annotations in queries. This view consists of two columns. The First column (left) lists the existing attributes while the second one (right) shows the associated attribute values. Templates[v] are ordinary Wiki pages that - according to the transclusion principle - are modules that can be integrated into other pages (for commonly used elements). They serve MediaWiki as tools used for example to create attributes of the same annotations. The syntax of the use of templates is *{{Template: page name}}*.

Figure 2: Developer view in SOFISwiki, displaying the internal representation of attributes

*Visualisation of the indicators*

To evaluate the indicators on the basis of the SOFISwiki structure we selected certain attributes in the developer view, values thereof serve as the basis for the visualisation of these indicators. Regarding the indicator "disciplinary area", the SOFISwiki field "main classification search" ("Hauptklassifikationsuch" in the developer's view) was queried and evaluated. For the remaining three features of "research activity", "type of funding" and "qualification" new attributes needed to be generated from existing ones based on templates and additional PHP extensions. For the visual representation of the indicators, the visualisation extensions available from Semantic MediaWiki were used and adapted accordingly (e.g. Sparkline, D3, jqPlot)[vi]. The implementation required technical programming adjustments based on templates.

The prototype concept assumes that the user wishes to make some selections on the project data corpus. Therefore, it is possible to reduce the visualisation to a certain status of projects (completed, starting or current) and geographical area (Germany or the complete corpus, i.e. Germany, Austria and Switzerland). This feature is supported by a self-developed PHP extension which is set up as a special site. As the dataset used in MoBi exclusively contains completed projects, only these were considered by the implementation (see Figure 3).

Figure 3: Project selection criteria by project status and geographical area

After selecting project status and geographical area users can select the indicator to be visualised (Figure 4, in descending order: research activity, discipline area, type of funding, qualification).

Figure 4: Selection list of indicators and time slice

Furthermore, users should be able to determine a time slice (selbstdefinierter Zeitraum) or to view the default time period (Gesamtzeitraum) (Figure 4). Therefore, we developed a PHP extension allowing the selection of years to be used for the chosen indicators (Figure 5).

Figure 5: Selection by year

## Research Activity

The "research activity" indicator informs about the number of projects per year, defined as the number of completed projects in MoBi. The SOFISwiki field "Year End" (Jahrgang Ende) was hence generated with a Wiki template. The respective value of the attribute is extracted from a SOFISwiki field, "duration until" (Laufzeit von) by a template and a respective figure is inserted into the newly created field. Diverse outcome formats of the Semantic MediaWiki were tested for the representation of this indicator, e.g. Sparkline and jqPlot. Figure 6 shows the result of a jqPlot[vii] presentation in a bar chart.

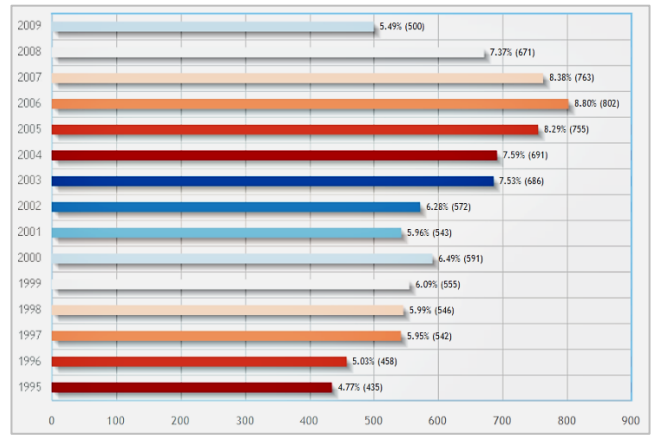

Figure 6: Bar chart visualisation of research activity (number of completed projects per year)

The project scope was highest in 2006; i.e. in 2006 the highest annual number of educational research projects was carried out looking at the period from 1995 to 2009. The share of the projects changes over time and is not linear.

## Disciplinary area

To provide a disciplinary distribution of the field the indicator "disciplinary area" relates all research projects to one of 12 areas based on the Social Sciences classification[viii] and is determined from the SOFISwiki

attribute "main classification search" (Hauptklassifikation-such). These are:

- Social Sciences and Humanities
- Sociology
- Population Science
- Political Science
- Education
- Psychology
- Communication Sciences
- Economics
- Social Policy
- Labour market and occupational research
- Interdisciplinary Subjects
- History

To visualise the indicator "disciplinary area" we used the following display options provided by Semantic MediaWiki: Sparkline, jqPlot, D3 and Tag cloud. For example, Figure 7 displays disciplines as Tag clouds[ix].

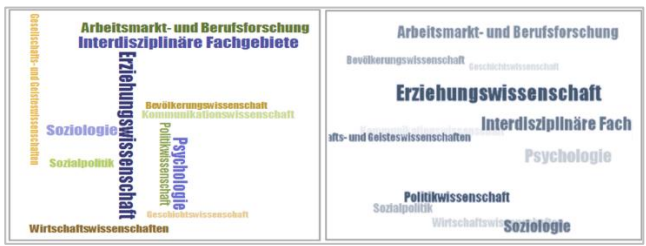

Figure 7: Tag cloud visualisation of discipline distribution

The font size indicates how strongly a discipline was represented in the period from 1995 to 2009. In this time period, education (Erziehungswissenschaft) is the most frequent discipline, followed by psychology.

*Type of funding*

To investigate influences of the funding and financing of the projects, a distinction is drawn between institutional (in-house) projects, third-party funded research and contract research. The type of funding is determined from the SOFISwiki attribute "type of research" (Forschungsart). In SOFISwiki the attribute can include nine possible features or their combinations:

- Contract research
- Third-party funded research
- In-house project
- Expertise
- Doctoral project
- Habilitation project
- Other exam thesis
- Other
- Unspecified

By using the self-developed PHP extension and the Wiki template, the new field "type of funding" was generated by extracting the rate of each of the three features "in-house projects" (Eigenprojekt), "third-party funded research" (Gefördert) and "contract research" (Auftragsforschung) from the nine possible features. The extensions $D3^x$ and jqPlot served to generate the results as shown in Figure 8 (D3: bubble chart and treemap, JqPlot: pie and donut).

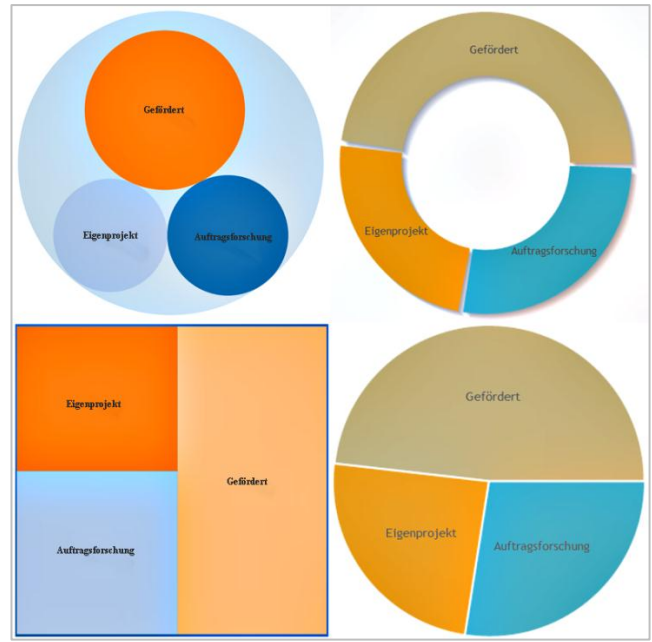

Figure 8: Different ways of visualising type of funding

To visualise the indicator over time, the jqPlot series[xi] extension was applied to the processed funding types. Results are shown in Figure 9.

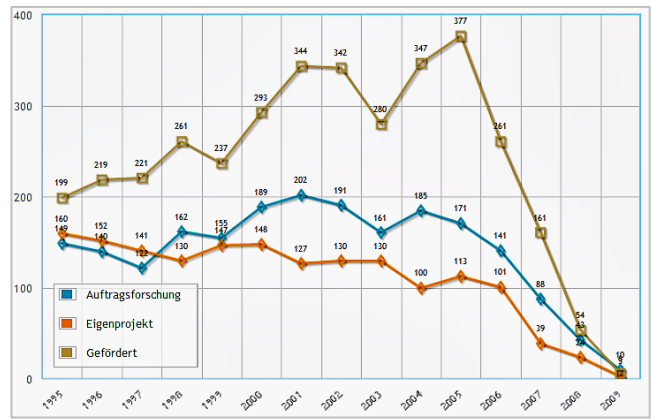

Figure 9: Number of different types of funding per year

Between 1995 and 2009, the number of third-party funded projects far exceeded the number of projects assigned to the other two funding types. Over time, a continuous decrease of institutionally (in-house) funded projects is observable, while since 1997 the proportion of contract research has been rising. From 2005 on, a significant decrease is evident for all types of funding.

Because it can be assigned to more than one funding type, one and the same project might be defined as an "in-house project" as well as a "third-party funded" or "contract research". It is thus impossible to allocate projects to just one type of funding, therefore, we only present absolute figures and refrain from calculating relative figures.

*Qualification*

The qualification of young scientists is an important indicator in the evaluation of research organisations and is based on the number of completed theses. To distinguish between doctoral and habilitation theses the indicator "qualification" was introduced. These values are programmatically read from the SOFISwiki field "type of research" (Forschungsart) and are inserted in the newly generated field "qualification". In its visual presentation, the same visualisation extensions were used as for "type of funding". Figure 10 shows how the proportion of doctoral and habilitation theses changed in the course of time from 1995 to 2009.

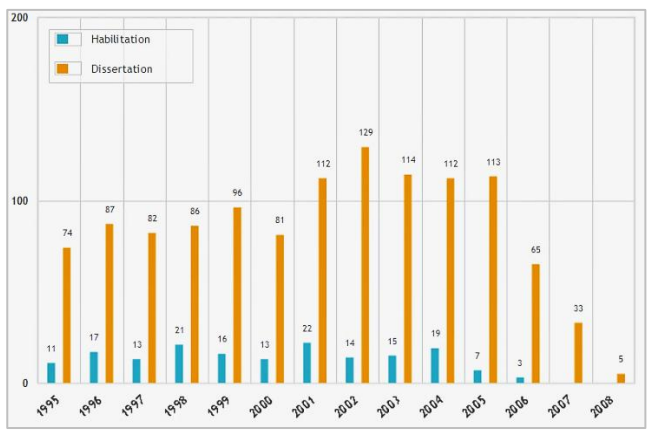

Figure 10: Bar chart visualisation of qualification theses per year

Overall, the results of the analyses show a strong increase of doctoral theses and a decrease of habilitation theses.

## Conclusions and discussion

Visualisation of information is challenged by the requirement of effectively presenting informational content and giving users optimal access to information. This can be enhanced by using colours and structural elements, thus, the human-computer interaction can be improved. For each parameter, the type of visualisation must be chosen in a way so that specific informational content is adequately presented.

Semantic MediaWiki is a powerful software for the visualisation of various data sets, yet, it is not sufficiently flexible to meet the specific requirements required by MoBi. We therefore had to expand functionalities by introducing our own scripts and templates to reach the desired results. In the MoBi prototype, difficulties emerge from the fact that some indicators are assigned to more than one value: the existence of different counting models poses specific demands as to the assignment of one respectively more than one value.

Other critical issues concern the idiosyncratic database from which the corpus of educational research projects was extracted. To our knowledge, no international database exists that would be comparable to the content area and metadata structure of SOFISwiki. It is thus impossible to draw a comparison based on comparative external data.

In a next step, the prototype will be expanded by lifting the limitation imposed by the restricted corpus for the MoBi project and including the entire SOFISwiki corpus. We will include projects that are still in their beginning and current projects as well as other geographical areas. Visualisation of significant deviations or anomalies across time is targeted as well.

## ACKNOWLEDGEMENTS

We would like to express our grateful thanks to Gwen Schulte (DIPF) and Julia Achenbach (GESIS) for their help with the translation and careful proofreading.

## Curriculum Vitae

Karima Haddou ou Moussa is software developer at GESIS in the department "Knowledge Technologies for the Social Sciences (WTS)" and a Master Degree student at the University of Applied Sciences Bonn-Rhein-Sieg.

Ute Sondergeld is information specialist and academic staff member at the Information Center for Education at DIPF.

Dr. Philipp Mayr is postdoctoral researcher and team head at GESIS department WTS. He is also a senior lecturer at Cologne University of Applied Sciences.

Peter Mutschke is Acting Head of the department WTS at GESIS and member of the executive committee of the Leibniz Research Alliance "Science 2.0".

Prof. Dr. Marc Rittberger is Director of the department Information Center for Education and Professor for Information Management at DIPF and the University of Applied Sciences Darmstadt. He is Deputive Executive Director of DIPF.

---

i http://sofis.gesis.org/sofiswiki/Hauptseite
ii http://semantic-mediawiki.org/
iii http://semantic-mediawiki.org/wiki/Help:Editing
iv http://semantic-mediawiki.org/wiki/Help:Properties_and_types
v http://semantic-mediawiki.org/wiki/Help:Semantic_templates
vi http://semantic-mediawiki.org/wiki/Help:Result_formats
vii http://semantic-mediawiki.org/wiki/Help:Jqplotchart_format
viii http://www.gesis.org/fileadmin/upload/dienstleistung/tools_standards/Kassifikation_Sozialwissenschaften_Stand_Juli_2013_dt_en__2_.pdf
ix http://semantic-mediawiki.org/wiki/Help:Tagcloud_format
x http://semantic-mediawiki.org/wiki/Help:D3_chart_format
xi http://semantic-mediawiki.org/wiki/Help:Jqplotseries_format